\documentclass[preprint, prc,nofootinbib,superscriptaddress,showpacs]{revtex4}

\usepackage{graphicx}
\usepackage{bm}
\usepackage{natbib}
\usepackage{amsfonts} 
\usepackage{amssymb}
\usepackage{bbm}
\usepackage{epsfig}
 \usepackage{epstopdf}
 \usepackage{amsmath,amssymb}
\usepackage{graphics}
\usepackage{subfigure}
\usepackage{colordvi}
\usepackage{multirow}
\usepackage{color}

\newcommand{\twopartdef}[4]
{
	\left\{
		\begin{array}{ll}
			#1 & \mbox{if } #2 \\
			#3 & \mbox{if } #4
		\end{array}
	\right.
}

\begin{document}
\newcommand{\lith}{$^6$Li~}
\newcommand{\alphad}{$\alpha+d$~}
\newcommand{\vh}{$V_H$~}
\newcommand{\vt}{$V_T$~}
\newcommand{\vm}{$V_M$~}
\newcommand{\vd}{$V_D$~}
\newcommand{\vg}{$V_G$~}
\newcommand{\be}{\begin{equation}}
\newcommand{\ee}{\end{equation}}
\newcommand{\bea}{\begin{eqnarray}}
\newcommand{\eea}{\end{eqnarray}}
\newcommand{\apjl}{Astrophys. J. Lett.}
\newcommand{\apjs}{Astrophys. J. Suppl. Ser.}
\newcommand{\aap}{Astron. \& Astrophys.}
\newcommand{\aj}{Astron. J.}
\newcommand{\araa}{Ann. Rev. Astron. Astrophys. } 
\newcommand{\mnras}{Mon. Not. R. Astron. Soc.}
\newcommand{\physrep}{Phys. Rept.}
\newcommand{\jcap}{JCAP}
\newcommand{\NP}{{\it Nucl. Phys.\,}}
\newcommand{\PR}{{\it Phys. Rev.\,}}
\newcommand{\etal}{{\it et al.}}

\begin{minipage}[t]{6.9in}
\end{minipage}

\title{The $\alpha + d \rightarrow ~ ^6\mathrm{Li} + \gamma $ astrophysical
$S$-factor and its implications for Big Bang Nucleosynthesis}

\author{A.\ Grassi}
\affiliation{Department of Physics, University of Pisa, Largo B. Pontecorvo 3, I-56127 Pisa, Italy}

\author{G.\ Mangano}
\affiliation{INFN, Sezione di Napoli, Complesso Univ. Monte S. Angelo, I-80126 Napoli, Italy}

\author{L.\ E.\ Marcucci}
\affiliation{Department of Physics, University of Pisa, Largo B. Pontecorvo 3, I-56127 Pisa, Italy}
\affiliation{INFN-Pisa, Largo B. Pontecorvo 3, I-56127 Pisa, I-56127 Pisa, Italy}

\author{O.\ Pisanti}
\affiliation{INFN, Sezione di Napoli, Complesso Univ. Monte S. Angelo, I-80126 Napoli, Italy}
\affiliation{Dipartimento di Fisica {\it Ettore Pancini}, Universit\`a di Napoli Federico II, Complesso Univ. Monte S. Angelo, I-80126 Napoli, Italy}

\date{\today}

\begin{abstract}
  The $\alpha+d\rightarrow\, ^6{\rm Li}+\gamma$ radiative capture is studied
  in order to predict the $^6$Li primordial abundance.
  Within a two-body framework, the $\alpha$ particle and the deuteron
  are considered the structureless constituents of \lith.
  Five $\alpha+d$ potentials are used to solve the two-body problem: four of
  them are taken from the literature, only one having 
  also a tensor component. A fifth model is here constructed
  in order to reproduce,
  besides the \lith static properties as binding energy, magnetic dipole
  and electric quadrupole moments, also the $S$-state asymptotic
  normalization coefficient (ANC). The two-body bound and scattering
  problem is solved with different techniques, in order to minimize
  the numerical uncertainty of the present results. The long-wavelength
  approximation is used, and therefore only the electric dipole and
  quadrupole operators are retained. The astrophysical $S$-factor is found
  to be significantly sensitive to the ANC, but in all the cases in good
  agreement with the available experimental data. The theoretical uncertainty
  has been estimated of the order of few \% when the potentials which
  reproduce the ANC are considered, but increases up to $\simeq 20$\%
  when all the five potential models are retained. The effect of this
  $S$-factor prediction on the \lith primordial abundance is studied,
  using the public code PArthENoPE. For the five models considered here
  we find $^6{\rm Li}/$H$ = (0.9 - 1.8) \times 10^{-14}$, with the baryon
  density parameter in the 3-$\sigma$ range of Planck 2015 analysis,
  $\Omega_b h^2= 0.02226 \pm 0.00023$.
\end{abstract}


\maketitle


\section{Introduction}
\label{sec:intro}

The $\alpha+d$ radiative capture
\begin{equation}
  \alpha+d \rightarrow ^6{\rm Li} +\gamma
  \label{eq:alphad}
\end{equation}
has recently received quite some interest, triggered by the so-called
\lith problem. In the theory of Big Bang Nucleosynthesis (BBN), even if it is a weak electric quadrupole transition, this reaction is important as represents the main $^6$Li production process. In 2006 Asplund {\it et al.} performed high
resolution observations of Li absorption lines in old halo stars~\cite{asp06}.
The $^6$Li/$^7$Li ratio was found to be of about $5\times 10^{-2}$,
more than two orders of magnitude larger than the expected BBN prediction. Since the analysis is performed on old stars, the quantity of
the present \lith should be a good estimate of the one
at the star formation, i.e.\ the same after BBN. This great discrepancy
is the so-called second Lithium problem. However, recent analyses with
three-dimensional modelling of stellar atmosphere, which do not assume local
thermodynamical equilibrium and include surface 
convection effects, show that these can explain 
the observed line asymmetry. The $^6\rm{Li}$ problem, therefore,
would be weakened~\cite{cayrel,perez,steffen,lind}. 

We recall that the BBN relevant energy window  is located
between 50 and 400 keV, and experimental studies of Eq.~(\ref{eq:alphad})
at these energies are very difficult, due to the exponential drop of the reaction
cross section as a consequence of the Coulomb barrier. Furthermore, this reaction is
affected by the
isotopic suppression of the electric dipole operator, as it will be discussed
in Sec.~\ref{subsec:transop}. The reaction~(\ref{eq:alphad}) was first studied
experimentally in the early 1980s~\cite{rob81}
and then thorough the 1990s~\cite{kie91,moh94,cec96,iga00}.
However the data in the BBN energy range were affected by large
uncertainties. The latest measurement is that performed
by the LUNA Collaboration~\cite{and14,tre17}. 

The theoretical study of this reaction is also very difficult, since,
in principle, we
should solve a six-body problem, i.e. we should consider the six nucleons
contained in the  $\alpha+d$ and \lith particles, and their interaction
with the photon. Such an approach
is known as the {\it ab-initio} method, and it has been used
only by Nollett {\it et al.} in Ref.~\cite{nol01}.
However, the numerical techniques used in Ref.~\cite{nol01} to solve the
six-body problem, i.e.\ the
variational Monte Carlo method, 
provide solutions for the initial and final state wave functions
with uncertainties at
the 10-20\% level. Since {\it ab-initio} methods are still nowadays
hardly implemented for $A>4$ radiative captures, the study of the reaction has
been done using a simplified model, where \lith is seen as
an $\alpha+d$ system and the problem is reduced to a two-body
problem. Then a crucial input for the calculation is represented
by the potential model, which describes the $\alpha+d$ interaction.
Five different potential models have been considered in this work,
four of them taken from
Refs.~\cite{ham10,tur15,muk11,dub98}, and a last one
constructed here starting
from the model of Ref.~\cite{dub98}, and then modifying it in order to
reproduce the asymptotic
normalization coefficient (ANC), i.e.\ the ratio between the $\alpha+d$
relative radial wave function in \lith and the Whittaker function
for large distances. It describes the bound-state wave function in
the asymptotic region. To be noticed that only the potential
of Ref.~\cite{dub98} and this last model
have a tensor component, necessary to describe the experimental values
for the \lith magnetic dipole and electric quadrupole moments. 
Our calculations have been performed using two methods to solve the two-body
Schr\"odinger equation, both for the bound and the scattering states,
in order to verify that our results are
not affected by significant numerical uncertainties.

The paper is organized as follows: in Sec.~\ref{sec:s-factor} we introduce
all the main ingredients of the present calculation for the astrophysical
$S$-factor and we present in Sec.~\ref{subsec:res}
our results. In Sec.~\ref{sec:li6PA} we discuss the implications
of the present calculated $S$-factor for the BBN prediction of
\lith abundance. We give our final remarks in Sec.~\ref{sec:concl}.

\section{The $\alpha + d $ \textit{S}-factor}
\label{sec:s-factor}

The $\alpha+d$ astrophysical $S$-factor $S(E)$, $E$ being the initial
center-of-mass energy, is defined as
\begin{equation}
S(E) = E\sigma(E)\exp(2\pi\eta)\:,
\label{eq:sfactor}
\end{equation}
where $\sigma(E)$ is the capture cross section, and
$\eta=2\alpha/v_{\rm rel}$ is the Sommerfeld parameter, $\alpha$ being the
fine structure constant and $v_{\rm rel}$ the $\alpha+d$ relative velocity.
With this definition, the $S$-factor has a smooth dependence on $E$
and can be easily extrapolated at low energies of
astrophysical interest.
The reaction cross section $\sigma(E)$ is given by
\begin{equation}
  \sigma(E)=\int {\mathrm{d}}\Omega_{\hat{\bf q}}
    \frac{{\mathrm{d}}\sigma}{{\mathrm{d}}\Omega_{\hat{\bf q}}}\:,
  \label{eq:sigma}
\end{equation}
where the differential cross section
${\mathrm{d}}\sigma/{\mathrm{d}}\Omega_{\hat{\bf q}}$ can be written as
\begin{equation}
\frac{\mathrm{d}\sigma}{\mathrm{d} \Omega_{\hat{\bf{q}}}} = 
\frac{e^2}{24 \pi^2 v_{rel}}\frac{q}{1+q/m_6}
\sum_{M_i \lambda M}\left| \hat{\epsilon}^{\dagger \lambda}_\mathbf{q}\cdot
\left\langle \Psi_{^6\mathrm{Li}}(M) | \mathbf{J}^\dagger(\mathbf{q})|
\Psi_{\alpha d} (M_i)
\right\rangle \right|^2 \:.
 \label{eq:crosssection}
\end{equation}
Here $m_6$ is the \lith mass, ${\bf q}$ is the photon momentum and
$\hat{\epsilon}^{\dagger \lambda}_\mathbf{q}$ its polarization vector,
$\mathbf{J}^\dagger(\mathbf{q})$ is the Fourier transform
of the nuclear electromagnetic current,
and $\Psi_{\alpha d}(M_i)$ and $\Psi_{^6\mathrm{Li}}(M)$ are the initial $\alpha+d$
and final \lith wave functions, with spin projection $M_i$ and $M$.
In Eq.~(\ref{eq:crosssection}),
we have averaged over the initial spin projections and summed
over the final ones.

In order to calculate the $\alpha+d$ cross section, it is necessary to
evaluate the \lith and $\alpha+d$ wave functions. This point is
described in the next Subsection.

\subsection{The $^6$Li and $\alpha+d$ systems}
\label{subsec:li6-ad}

A crucial input for our calculation is represented by the \lith
and $\alpha+d$ wave functions. We consider first the bound state.
The nucleus of \lith has $J^\pi=1^+$, a binding energy $B$
respect to the $\alpha+d$ threshold of 1.475 MeV~\cite{dub98}, a
non-null electric quadrupole moment
$Q_6=-0.0644(7)$ fm$^2$ and a
magnetic dipole moment $\mu_6=-0.822$ $\mu_N$~\cite{dub98}.  
As it was shown in Ref.~\cite{muk11}, the astrophysical $S$-factor
at low energies is highly sensitive not only to the \lith binding energy
$B$ and the \alphad scattering phase shifts, but also to the \lith
$S$-state asymptotic normalization coefficient (ANC).
This quantity is crucial due to the peripheral nature
of the $\alpha+d$ reaction at low energies, where only the tail of
the \lith wave function gives most of the contribution in the
matrix element of Eq.~(\ref{eq:crosssection}). 
The $S$-state ANC is defined as
\begin{equation}
  C_{\ell=0} =
  \lim_{r\rightarrow+\infty} \frac{\varphi(r)}{W_{-\eta,\ell+1/2}(r)}\bigg|_{\ell=0}
  \:,
\label{eq:anc0}
\end{equation}
where $\varphi(r)$ is the $S$-state \lith reduced wave function,
$W_{-\eta,\ell+1/2}(r)$ is the Whittaker function, $\eta$ is the
Sommerfeld factor and $\ell=0$ for the $S$-state ANC.
Its experimental value for \lith is ANC$_{\rm exp}=(2.30\pm 0.12)$
fm$^{1/2}$~\cite{tur15}.

In the present study we consider the \lith nucleus as a compound system, made
of an $\alpha$ particle and a deuteron. In fact, as it was shown in
Ref.~\cite{dub94}, the
$\alpha+d$ clusterization percentage in \lith can be up to about 60-80\%.
Therefore, we solve in this work a two-body problem, including both
$S-$ and $D$-states in the $\alpha+d$ bound system. The first observable that
we will try to reproduce is the binding energy, but we will consider
also the above mentioned observables of \lith.

At this point, an important input for the calculation is represented by the
$\alpha+d$ potential. The different models considered in this work will
be discussed below. 

\subsubsection{The $\alpha+d$ Potentials} 
\label{subsubsec:potentials}

For our calculation we consider five different potential models.
The use of so many models allows us to get a hint on the theoretical
uncertainty arising from the description of the \lith nucleus and
the $\alpha+d$ scattering system. Four of these potentials are taken
from Refs.~\cite{ham10,muk11,tur15,dub98}, while the last one has
been constructed in the present work as described below. The
physical constants present in each potential as listed on the original
references are summarized in Table~\ref{tab:dataauthor}.

\begin{table}[t]
\centering
\begin{tabular}{|c|c|c|c|}
\hline
~ ~ ~~~~~~~~~& ~~~~~~units~~~~~~ & $V_H$ and $V_M$ & $V_T$, $V_D$ and $V_{G}$\\ 
\hline \hline
$A_d$ &-& 2.01411 & 2 \\ 
$A_\alpha$ &-& 4.00260 & 4 \\ 
$m_u $ & MeV & 931.494043 & \underline{938.973881}\\ 
$\mu$ & MeV & \underline{1248.09137} & \underline{1251.96518} \\ 
$\hbar^2/2\mu$ & MeV fm$^2$ & \underline{15.5989911176} & 15.5507250000 \\ 
$\alpha$ &-& 7.297352568$\times$10$^3$ &  \underline{7.297405999$\times$10$^3$} \\ 
$\alpha \hbar c $ &MeV fm& \underline{1.4399644567}	& 1.4399750000\\ 
$B$ & MeV & 1.474 & 1.475 ($V_T$) \\
    &     &       & 1.4735 ($V_D$ and $V_G$)\\
\hline
\end{tabular}
\caption{Set of the constants present in the five $\alpha+d$
	potential models,
	labelled as $V_H$, $V_T$, $V_M$, $V_D$,
	taken from Refs.~\cite{ham10,tur15,muk11,dub98}, respectively,
        and $V_{G}$,
	constructed in the present work.
	$A_\alpha$ ($A_d$) is the mass numbers of the $\alpha$ ($d$) particle,
	$m_u$ is the mass unit, equal to the  atomic mass unit for $V_H$ and $V_M$,
	and to the average nucleon mass for $V_T$, $V_D$ and $V_{G}$,
	$\mu$ is the $\alpha+d$ reduced mass, $\alpha$ is the fine-structure
	constant and $B$ is the \lith binding energy respect to the
	$\alpha+d$ threshold. The underlined quantities are deduced from other
	data given by the authors
        in the original
        references~\cite{ham10,tur15,muk11,dub98}.\vspace{0.2 cm}}
\label{tab:dataauthor}
\end{table}

The first potential used in our study has been taken from Ref.~\cite{ham10}
and has the form
\begin{equation}
V_H(r) = -V_C^{\ell} \left[1+\exp\left(\frac{r-r_0}{a}\right)\right]^{-1}
+ V_{SO} \frac{\lambda^2 \mathbf{L}\cdot \mathbf{S}}{r}
\frac{\mathrm{d}}{\mathrm{d}r}
\left[1+\exp\left(\frac{r-r_0}{a}\right)\right]^{-1} + V_{Coul}^{(m)}(r)\:.
\label{eq:potHam}
\end{equation}
It contains a spin-independent Wood-Saxon component, a spin-orbit interaction
term, and a modified Coulomb potential, which is written as
\begin{equation}
  V_{Coul}^{(m)}(r)=Z_\alpha Z_d \:\alpha \:\twopartdef
  { \left[3-\left(r/r_0\right)^2 \right]/\left(2 r_0\right) } {r \le r_0} {1/r}
  {r > r_0}\:.
  \label{eq:vcmod}
\end{equation}
The values for all the parameters present in Eqs.~(\ref{eq:potHam})
and~(\ref{eq:vcmod}), as well as those of the following
potentials, are listed in Table~\ref{tab:potentialConstants},
apart from $r_0$, which is $r_0=1.25\,A^{1/3}$ fm, with $A=6$.
To be noticed that this potential does not reproduce the experimental
value of the ANC, as it has been noticed in Ref.~\cite{muk11}, and we
have ourselves verified by calculating the \lith properties (see below).

The second potential is taken from Ref.~\cite{tur15},
and can be written as
\begin{equation}
V_T(r) = -V_0^{\ell} \exp\left(-\frac{r^2}{a_\ell^2}\right) + V_{Coul}(r)\:.
\label{eq:potTur}
\end{equation}
It is therefore the sum of a Gaussian function and a Coulomb point-like
interaction $V_{Coul}(r)=Z_\alpha Z_d \:\alpha/r$. It reproduces
the experimental ANC for the $^6$Li (see below).

The third potential is obtained by adding to the $V_H$ potential
of Ref.~\cite{ham10}, a new term $V_N(r)$, such that the new potential
\begin{equation}
  V_M(r)=V_H(r)+V_N(r)
  \label{eq:potMuk}
\end{equation}
reproduces the experimental ANC~\cite{muk11}. 
The procedure to obtain $V_N(r)$ is discussed at length in Ref.~\cite{muk11}.
Here we have generalized it to the coupled-channel case,
and it will be discussed below.

The potentials $V_H$, $V_T$ and $V_M$ considered so far are
central potentials, which have, at
maximum, a spin-orbit term. Therefore, these potentials are unable to
give rise to the $^3D_1$ component in the \lith wave function. The
non-zero \lith quadrupole moment has induced us to consider also potentials
which include a tensor term. In this study, we have used the potential
of Ref.~\cite{dub98}, which can be written as
\begin{equation}
  V_D(r) = -V_0^{\ell J} \exp\left(-\frac{r^2}{a^2}\right) - V_1^{\ell}
  \exp\left(-\frac{r^2}{b^2}\right)\left[ 6\frac{(\mathbf{S}\cdot
      \mathbf{r})^2}{r^2} - 2 \mathbf{S^2}\right]+V_{Coul}(r)\:,
\label{eq:potDub}
\end{equation}
where $\mathbf S$ is the spin operator acting on $^6$Li.
The coefficients $V_0^{\ell J} \equiv V_0^{01}$ and $V_0^{\ell J} \equiv V_0^{21}$
have been taken from Ref.~\cite{dub98}. However, in Ref.~\cite{dub98}
this potential was used only for the bound-state problem.
Therefore, we have modified
the potential in order to reproduce also the scattering phase-shifts up to
$\ell=2$. In order to do so, the depth $V_0^{\ell J}$ has been
fitted to the experimental scattering phase-shifts
for every initial channel, minimizing the $\chi^2$  of the calculated
phase shifts with respect to the available experimental data taken from 
Refs.~\cite{jen83,mci67,gru75,bru82,kel70}. In this procedure,
we minimized the $\chi^2$ changing the value of $V_0^{\ell J}$.
We have used both the bisection and the Newton's method, finding
no difference between the calculated values of $V_0^{\ell J}$. These
have been listed in Table~\ref{tab:potentialConstants}.

As in the case of $V_H$, also the $V_D(r)$ potential,
does not reproduce the \lith ANC. Therefore,
we have constructed a new model
generalizing the procedure of Ref.~\cite{muk11} to the
coupled-channel case. 
We start from a generic Hamiltonian operator $H_0$, for which we know the
bound state radial eigenfunction $\vec \varphi(r)$, the corresponding binding
energy $B$ and the ANC $C_0$ for the $S$-state. We have defined
$\vec \varphi(r)$ to be the vector containing the $S$- and $D$-state bound
wave functions, i.e.\ $\vec \varphi(r) = (\varphi_0,~ \varphi_2)$
and normalized it to unity, i.e.\
\begin{equation}
  \int_0^\infty dx x^2 (\vec \varphi(x)\cdot \vec \varphi(x))=1
  \label{eq:norm}\ .
\end{equation}
We want
to find a potential part of an Hamiltonian which has the same binding energy,
but the correct value for $C_0$, which we will call $C_0^N$.
As an \textit{Ansatz}, we assume that our new solution has the form
\begin{equation}
  \vec \phi(r) = \vec \varphi(r)/\gamma(r)
  \label{eq:verphi}\ ,
\end{equation}
with
\begin{equation}
  \gamma (r ) \equiv \tau^{-1/2}\:\left[1+(\tau-1)\int_0^r\:\mathrm{d}x\:
    x^2(\vec \varphi(x)\cdot \vec \varphi(x))\right]\:,
  \label{eq:gamma}
\end{equation}
where $\tau$ is a parameter to be fitted to the experimental ANC value. 
This solution is correctly normalized and the new ANC $C_0^N$ is given by
\begin{equation}
C_0^N=
 \lim_{r\rightarrow+\infty} \frac{ \phi_0(r)}{W_{-\eta,1/2}(r)}=
  \frac{1}{\sqrt{\tau}}\lim_{r\rightarrow +\infty}\frac{ \varphi_0(r)}{W_{-\eta,1/2}(r)}=\frac{C_0}{\sqrt{\tau}}\:.
\end{equation} 
It is then enough to choose 
$\tau=(C_0/C_0^{exp})^2$, so that $C_0^N=C_0^{exp}$.
For the $V_M$ potential, $\tau=1.378$~\cite{muk11},
while for this coupled-channel case $\tau=1.181$.

In order to obtain the new wave function $\vec \phi(r)$, we define
a new Hamiltonian operator as
\begin{equation}
  H = H_0 + V_N\:,
  \label{eq:hnew}
\end{equation}
and we impose
\begin{equation}\label{appeq:h}
H\vec \phi(r) = -B\vec \phi(r)\:,
\end{equation}
knowing that 
\begin{equation}\label{appeq:h0}
H_0\vec \varphi(r) = - B \vec \varphi(r)\:.
\end{equation}
Subtracting Eq.~(\ref{appeq:h0}) from Eq.~(\ref{appeq:h}), we obtain
\begin{equation}
  \frac{\hbar^2}{2\:\mu}\left\{\left[-2\left(\frac{\gamma'(r)}
    {\gamma(r)}\right)^2 
    +\frac{\gamma''(r)}{\gamma(r)}\right]\vec\varphi(r)+2
  \frac{\gamma'(r)}{\gamma(r)}\vec\varphi\:'(r)\right\}+
  V_N(r) \vec\varphi(r)=0\:,
\end{equation}
which can be re-written as 
\begin{align}
  \nonumber V_N(r)\vec\varphi(r) &= -\frac{\hbar^2}{2\:\mu}\left\{\left[
    -2\left(\frac{\gamma'(r)}{\gamma(r)}\right)^2 
    +\frac{\gamma''(r)}{\gamma(r)}\right]\vec\varphi(r)+2
  \frac{\gamma'(r)}{\gamma(r)}\vec\varphi\:'(r)\right\}\\
  \nonumber&= -\frac{\hbar^2}{2\:\mu}\left\{2\left[-\left(
    \frac{\gamma'(r)}{\gamma(r)}\right)^2 
    +\frac{\gamma''(r)}{\gamma(r)}\right]\vec\varphi(r)+2
  \frac{\gamma'(r)}{\gamma(r)}\vec\varphi\:'(r)-
  \frac{\gamma''(r)}{\gamma(r)}\vec\varphi(r)\right\}\\
&= -\frac{\hbar^2}{2\:\mu}\left\{2\left[
    \frac{\mathrm{d}^2}{\mathrm{d}r^2}\log\gamma(r)\right]\vec\varphi(r)+2
  \frac{\gamma'(r)}{\gamma(r)}\vec\varphi\:'(r)-
  \frac{\gamma''(r)}{\gamma(r)}\vec\varphi(r)\right\}\:.
\end{align}
Writing explicitly $\vec \varphi(r)$ and $\gamma(r)$, and
assuming for simplicity $V_N(r)$ to be diagonal, we get
\begin{align}
[V_N(r)]_{11}& = -2\frac{\hbar^2}{2\:\mu}\bigg\{
\frac{\mathrm{d}^2}{\mathrm{d}r^2}\ln\gamma(r) + \frac{\tau-1}{\gamma(r)}
\varphi_2^2(r)\frac{\mathrm{d}}{\mathrm{d}r}\ln
\frac{\varphi_0(r)}{\varphi_2(r)}\bigg\}\:,
\label{newp1}
\\
[V_N(r)]_{22}& = -2\frac{\hbar^2}{2\:\mu}\bigg\{
\frac{\mathrm{d}^2}{\mathrm{d}r^2}\ln\gamma(r) - \frac{\tau-1}{\gamma(r)}
\varphi_0^2(r)\frac{\mathrm{d}}{\mathrm{d}r}\ln
\frac{\varphi_0(r)}{\varphi_2(r)}\bigg\}\:.
\label{newp2}
\end{align}
Note that if we consider only central potentials, then
the potential $V_N(r)$ acts only on the \lith $^3$S$_1$ state, and reduces to
\begin{equation}
  V_N(r) = -2 \frac{\hbar^2}{2\mu} \frac{\mathrm{d}^2}{\mathrm{d}r^2}
  \ln\left[1+(\tau-1)\int_0^r~\varphi^2(x) \mathrm{d}x\right]\:,
\label{eq:potMukN}
\end{equation}
as obtained in Ref.~\cite{muk11}. This is the term added in
Eq.~(\ref{eq:potMuk}).
As we have seen, this new term should give rise to no changes in the
binding energy, nor
in the scattering phase-shifts with respect to those
evaluated with the $H_0$. This has been verified 
with a direct calculation.

Finally, this last potential is defined as
\begin{equation}
  V_G(r)=V_D(r)+V_N(r)\ ,
  \label{eq:potGr}
\end{equation}
where $V_N(r)$ is given in Eqs.~(\ref{newp1}) and~(\ref{newp2}). 

\begin{table}[t!]
\centering
\begin{tabular}{|c|c|c|c|c|c|c|cccccccc|}
\hline
\multicolumn{1}{|l|}{{~Potential~}} & \multicolumn{14}{c|}{{Parameters}} \\
\hline
\multirow{2}{*}{\begin{tabular}[c]{@{}c@{}}$V_H$ \& $V_M$\end{tabular}}
& $V_c$  & $V_c^{\ell\neq 0}$ & $V_{SO}$   & $R$                   & $\lambda$
& $a$    &                             &                       &
&        &                             &                       &
&      \\
& 60.712 & 56.7    & 2.4        & 2.271 & 2          & 0.65       &
&        &         &            &       &            &            &      \\
\hline
\multirow{2}{*}{$V_T$}
& $V_0$  & $a_0$      & $V_0^{10}$ & $a_1$    & $V_0^{11}$ & $a_1$        &
\multicolumn{1}{c|}{$V_0^{12}$} & \multicolumn{1}{c|}{$a_2$}        &
\multicolumn{1}{c|}{$V_0^{21}$} & \multicolumn{1}{c|}{$a_2$}  &
\multicolumn{1}{c|}{$V_0^{22}$} & \multicolumn{1}{c|}{$a_2$}  &
\multicolumn{1}{c|}{$V_0^{23}$} & $a_2$  \\
& 92.44  & 0.25               & 68.0       & 0.22                  & 79.0
& 0.22       & \multicolumn{1}{c|}{85.0}       & \multicolumn{1}{c|}{0.22}
& \multicolumn{1}{c|}{63.0}       & \multicolumn{1}{c|}{0.19}
& \multicolumn{1}{c|}{69.0}       & \multicolumn{1}{c|}{0.19}
& \multicolumn{1}{c|}{80.88}      & 0.19 \\
\hline
\multirow{2}{*}{\begin{tabular}[c]{@{}c@{}}$V_D$ \& $V_G$\end{tabular}}
& $V_0$  & $a$ & $V_1$  & $b$  & $V_0^{10}$ & $V_0^{11}$
& \multicolumn{1}{c|}{$V_0^{12}$} & \multicolumn{1}{c|}{$V_0^{22}$}
& \multicolumn{1}{c|}{$V_0^{23}$} & & & & &      \\
& 71.979 & 0.2  & 27.0 & 1.12   & 77.4  & 73.08
& \multicolumn{1}{c|}{78.42}    & \multicolumn{1}{c|}{72.979}
& \multicolumn{1}{c|}{86.139}   &  &  & & &      \\ \hline
\end{tabular}
\caption{Parameters present in the five potential models used in this work.
	The parameters $V_c$, $V_{SO}$, $V_0$, $V_1$ and $V_0^{\ell J}$
	are given in MeV, all the others are in fm. We have used the notation
	$\ell$ for the orbital angular momentum and $J$ for the total angular
	momentum. $V_0$ and $V_c$ are used for the $\ell=0$ state.
	The Gaussian width $a_\ell$
        for the $V_T$ potential is written to the right
	of each potential depth.\vspace{0.2 cm}}
\label{tab:potentialConstants}
\end{table}

\subsubsection{Numerical methods}
\label{subsubsec:nummeth}

In order to solve the Schr\"odinger equation,
both for the initial and final states,
two methods have been adopted, the Numerov's  and the variational
method. In particular, we have used the Numerov's method for the
bound-state problem, and the variational one for both the bound- and
the scattering-state problem. 
The convergence of the two methods has been tested, proving that both
methods give the same numerical results for the $S$-factor,
with a very good accuracy. The choice of the variational method for the
scattering-state is related to the fact that this method is simpler to
be extended to the coupled-channel case.
In fact, the Numerov's method, even for the bound-state problem, needs some
improvement respect to the single-channel case. Here we have
proceeded as follows.
The reduced radial waves solutions for the $^3S_1$ ($\varphi_0$) and
$^3D_1$ ($\varphi_2$) states of \lith must satisfy the coupled
equations
\begin{eqnarray}
  \varphi_0''(r)+ \varphi_0(r)\:\frac{\hbar^2}{2\mu}[E-V_{00}(r)] & =&
  \varphi_2(r)\:\frac{\hbar^2}{2\mu} V_{02}(r) \label{eq:num1}\\
  \varphi_2''(r)+ \varphi_2(r)\:\frac{\hbar^2}{2\mu}[E-V_{22}(r)-
    \frac{12\mu}{\hbar^2 r^2}] & =& \varphi_0(r)\:\frac{\hbar^2}{2\mu}
  V_{20}(r)
\label{eq:num2}
\end{eqnarray}
We solve this system of equations iteratively. First we consider
$\varphi_0(r)$ to be zero. Eq.(\ref{eq:num2}) then becomes
\begin{equation}
  \varphi_2''(r)+ \varphi_2(r)\:\frac{\hbar^2}{2\mu}
         [E-V_{22}(r)- \frac{12\mu}{\hbar^2 r^2}]  = 0\:,
         \label{eq:varphi2}
\end{equation}
which we solve with the standard Numerov's algorithm, obtaining
$E\equiv E_2$ and $\varphi_2(r)$.
Then we calculate the solution of Eq.~(\ref{eq:num1})
giving an initial value for the normalization ratio $a$, defined as
\begin{equation}
  a=\lim_{r\rightarrow +\infty} \frac{\varphi_2(r)}{\varphi_0(r)}\ ,
  \label{eq:def_asynnorm}
\end{equation}
and applying again the Numerov's algorithm to obtain
$\varphi_0(r)$.
With the evaluated $\varphi_0(r)$, we calculate again
$\varphi_2(r)$ with Eq.~(\ref{eq:num2}), and so on, until we
converge both for $E$ and for $a$ within the required accuracy.

The method for the single-channel scattering problem is straightforward,
as the Numerov's outgoing solution from $r=0$ to the final grid point
is matched to the function
\begin{equation}
  \varphi(r)=\cos\delta_l F_l(\eta;kr)+\sin\delta_l G_l(\eta;kr)\ ,
  \label{eq:asymp-scatt}
\end{equation}
which is normalized to the unitary flux. Here $F_l(\eta;kr)$
and $G_l(\eta;kr)$ are the regular and irregular Coulomb functions,
and $k$ the $\alpha+d$ relative momentum. 
The scattering phase-shift $\delta_l$ is then easily obtained.

The variational method has been used for both the bound and
the scattering states. 
For the bound state, we expand the wave function as
\begin{equation}
\Psi(\mathbf{r})=\sum_{\alpha i}~c_{\alpha i}\:f_{\alpha i}(r)\:|\alpha\rangle\:,
\label{eq:varExp}
\end{equation}
where
$|\alpha\rangle \equiv \sum_{m\,\sigma}\: \langle \ell m S \sigma | J M \rangle
Y_{\ell m}(\hat{\mathbf{x}})\chi_{S\sigma}$, $f_{\alpha i}(r)$ are orthonormal
functions and $c_{\alpha i}$ are unknown coefficients. 
We use the Rayleigh-Ritz variational principle to
reduce the problem to an eigenvalue-eigenvector problem, which can be solved
with standard techniques (see Ref.~\cite{Mar11} for more details).
Here we use a basis function defined as
\begin{equation}
  f_{\alpha i}(r)=\sqrt{\frac{\gamma_\alpha i!}{(i+2)!}}~
  L_i^{(2)}(\gamma_\alpha r)~{\rm e}^{-\gamma_\alpha\:r/2}\:,
\label{eq:laguerrre}
\end{equation}
with $\gamma_\alpha=4$ fm$^{-1}$ for each $\alpha$,
and $L_i^{(2)}(\gamma_\alpha r)$ are Laguerre polynomials.
Note that so defined, these functions are orthonormal.

For the scattering problem the wave function is decomposed as
\begin{equation}
  \Psi(\mathbf{r}) = \Psi_c(\mathbf{r})+\frac{F_\ell(\eta;kr)}{kr}
  \:|\alpha\rangle +
  \sum_{\beta}\:{^J}R_{\alpha\beta}\:\frac{\tilde{G}_{\ell_\beta}(\eta; kr)}{kr}|
  \beta\rangle\:,
\label{eq:asymp-wf}
\end{equation}
where $^JR_{\alpha\beta}$ are unknown coefficients, $\Psi_c(\mathbf{r})$
has the same form as $\Psi(\mathbf{r})$ in Eq.~(\ref{eq:varExp}),
while ${F}_\ell(\eta;kr)$ and
$\tilde{G}_\ell(\eta;kr)=G_\ell(\eta;kr)(1-{\rm e}^{r/r_0})^{2\ell +1}$
are regular and (regularized for $r\rightarrow 0$) irregular Coulomb functions,
with $r_0$ a non-linear parameter of the order of 4 fm.
The Kohn variational principle is used to obtain
the unknown coefficients ${^J}R_{\alpha\alpha'}$ and $c_{\alpha i}$
of Eq.~(\ref{eq:varExp}), with a standard procedure as
outlined in Ref.~\cite{Mar11}.
\begin{figure}[t!]
\centering
\includegraphics[width=12 cm]{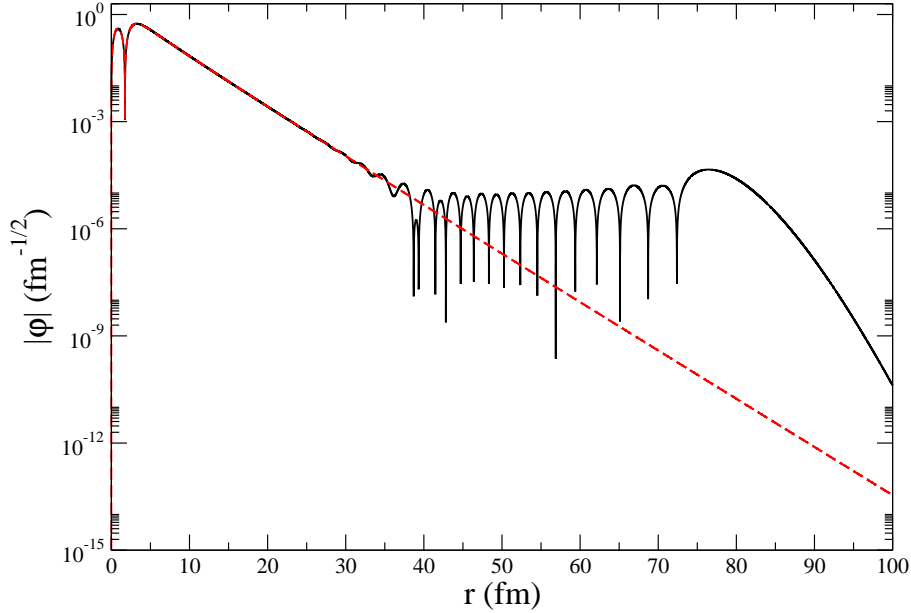}
\caption{The modulus for the $^6$Li wave function in logarithmic scale
  obtained with the variational (black) and the Numerov's (red dashed)
  methods using the $V_H$ potential.}
\label{fig:boundWaveVar}
\end{figure}

In Fig.~\ref{fig:boundWaveVar} we show a comparison for the \lith
reduced radial wave functions obtained, using the $V_H$ potential, with the
Numerov's or variational method. Similar results can be found for the
other potentials. As it can be seen by inspection of the figure,
the variational method is unable to reproduce the \lith wave function
at large distances, of the order of 30-40 fm. 
In this case the reduced wave function has been cured in order to get the
correct asymptotic behaviour. Within the Numerov's method,
the long range wave function is constructed by hand.
The agreement between the two methods is much nicer for the scattering
problem, although the Numerov's method has been
used for the single channels. In these cases, the
agreement between the two methods is at the order
of 0.1\%.

\subsubsection{The \lith nucleus and the $\alpha+d$ scattering state}
\label{subsubsec:results}

\begin{figure}
\centering
\includegraphics[width=1\textwidth]{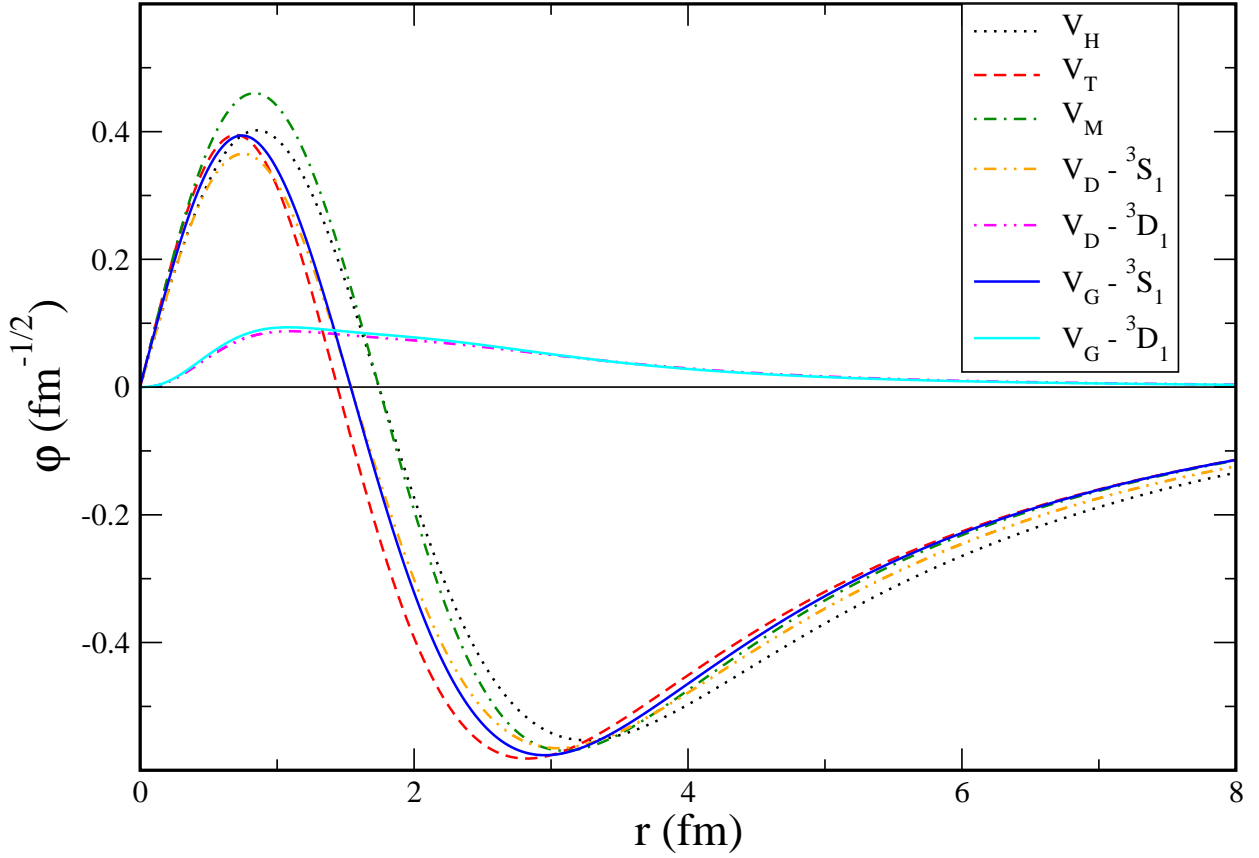}
\caption{The \lith reduced wave function evaluated with each potential model
  considered in this work.}
\label{fig:bndwaves}
\end{figure}

The \lith static properties, i.e.\ the binding energy respect to
the $\alpha+d$ threshold, the $S$-state ANC, the magnetic dipole moment
$\mu_6$ and the electric quadrupole moment $Q_6$ are given in
Table~\ref{tab:li6res}.
By inspection of the table we can conclude that each potential
nicely reproduces the experimental binding energy, while only
$V_T$, $V_M$ and $V_G$ give good values for the ANC. Also, the $V_D$ and
$V_G$ potentials are the only ones which include the $D$-state contributions
in the \lith wave function. Therefore, the values of
$\mu_6$ and $Q_6$ obtained with these potentials are closer to the
experimental values, while $\mu_6$ and $Q_6$ calculated with the
$V_H$, $V_T$, and $V_M$ potentials are simply those of
the deuterium.
Finally, we show in Fig.~\ref{fig:bndwaves} the \lith reduced wave function
evaluated with each potential.
The differences between the various potentials are quite pronounced
for $r\leq 6$ fm. However, this is not too relevant for
our reaction, which is peripheral and therefore most sensitive to the tail
of the wave function and to the $S$-state ANC. 
\begin{table}[t]
\begin{tabular}{|c|c|c|c|c|c|c|} \hline
$\qquad$&\vh & \vt & \vm & \vd & \vg & EXP.\\ 
 \hline
$B$ & 1.474 & 1.475 & 1.474 & 1.4735 & 1.4735 & 1.474 \\ 
$C_0$ & 2.70 & 2.31 & 2.30 & 2.50 & 2.30 & 2.30\\
$\mu_6$ & 0.857 & 0.857 & 0.857 & 0.848 & 0.848 & 0.822\\
$Q_6$ & 0.286 & 0.286 & 0.286 & -0.066 &-0.051 & -0.082\\
\hline
\end{tabular}
\caption{The \lith binding energy ($B$) in MeV, $S$-state ANC ($C_0$)
  in fm$^{1/2}$, magnetic dipole moment $\mu_6$ in $\mu_N$ and electric quadrupole moment $Q_6$ in fm$^2$
   are calculated with the five different
    potential models $V_H$, $V_T$, $V_M$, $V_D$, and $V_G$. The available
  experimental data are also shown.}
\label{tab:li6res}
\end{table}

\begin{figure}[tp]
\centering
\includegraphics[height=0.75\textheight]{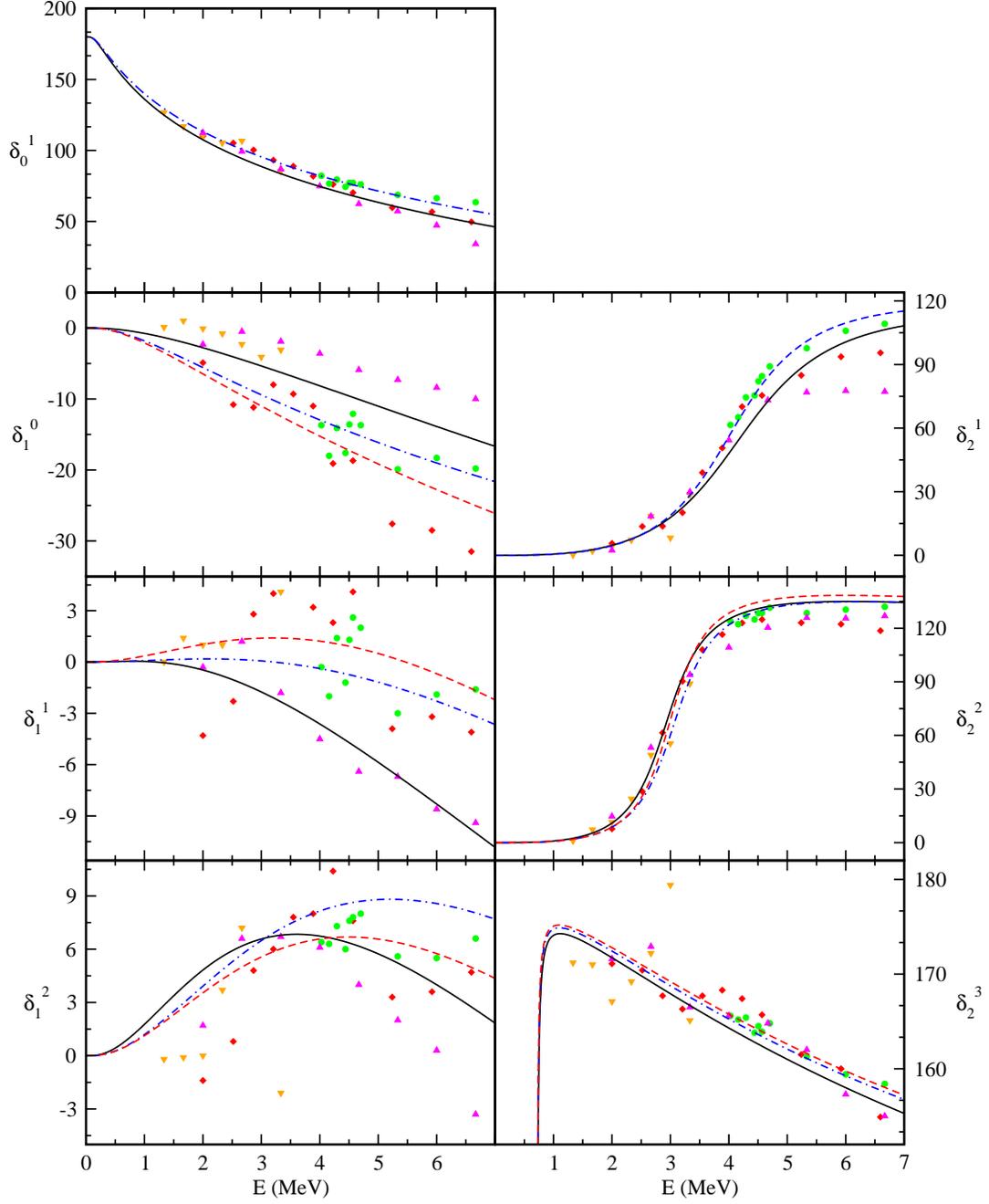}
\caption{The phase shift $\delta_\ell ^J$ for every partial wave $^3\ell_J$,
  where $\ell=\{0,~1,~2\}$. The data have been taken from
  Refs.~\cite{jen83,mci67,gru75,bru82,kel70}. 
  The phase shifts are given in degrees as a function of the
  center-of-mass relative energy in MeV.
  The shape of the experimental points indicates the article
  from which the data were taken: 
  we use circles~\cite{jen83}, triangles down~\cite{mci67},
  diamonds~\cite{gru75}, squares~\cite{bru82} and triangles up~\cite{kel70}.
  The calculated phase shifts are obtained with the $V_H$ and $V_M$
  (black solid line), $V_T$ (blue dash-dotted line),
  and $V_D$ and $V_{G}$ (red dashed line) potentials.}
\label{fig:scatteringresults}
\end{figure}

\begin{figure}[t!]
\includegraphics[width=1.\textwidth]{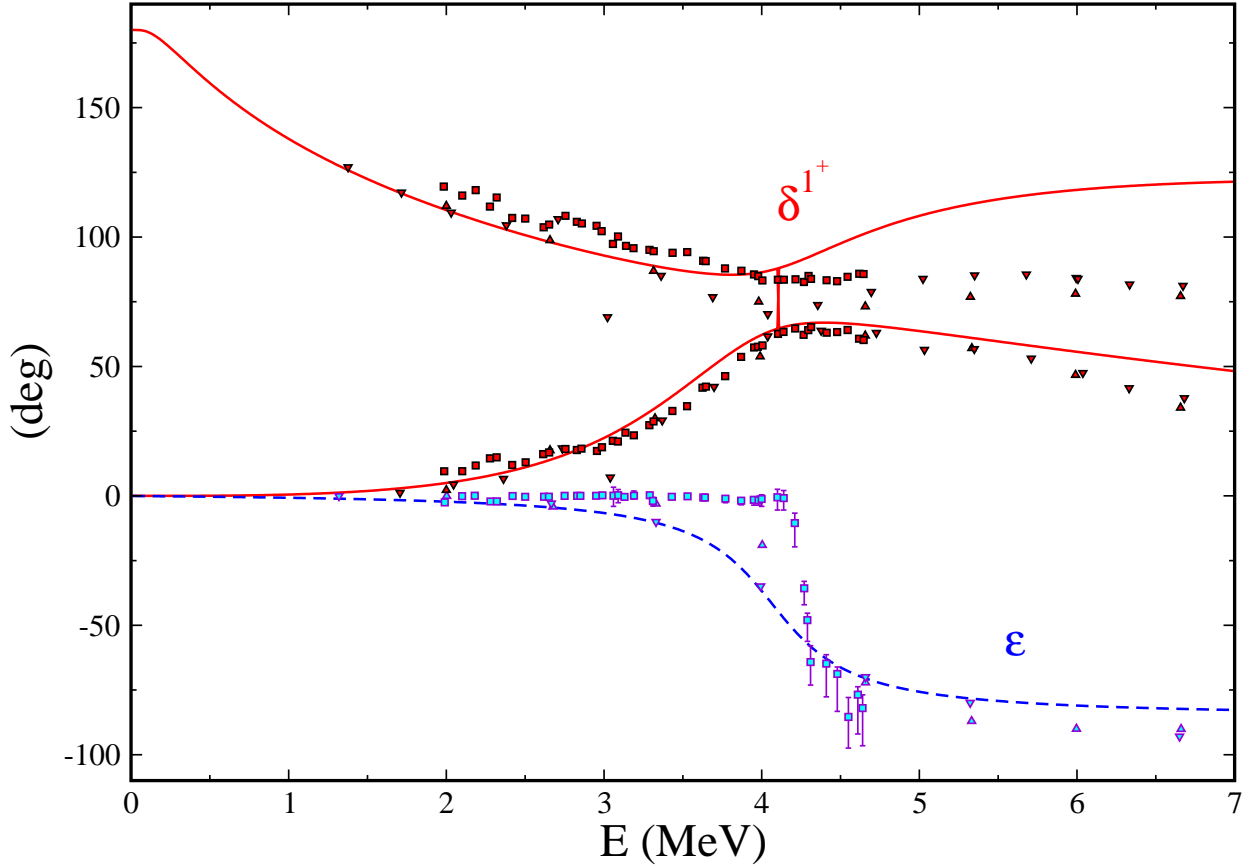}
\caption{Same as Fig.~\ref{fig:scatteringresults}, but for the coupled
  channels in the $J^\pi=1^+$ state. The phase shift results evaluated with
  the $V_D$ and the $V_{G}$ potentials for the coupled channels
  $^3S_1$ and $^3D_1$ are displayed with the two red solid lines,
  the results for the mixing angle $\varepsilon$ are displayed with the
  blue dashed line.}
\label{fig:scatteringresultsCC}
\end{figure}

For the initial $\alpha+d$ scattering state,
the scattering phase shifts obtained with each potential
are in good agreement with the experimental data, as it
can be seen 
in Fig.~\ref{fig:scatteringresults} for the single channels
and in Fig.~\ref{fig:scatteringresultsCC} for the coupled channels.
In particular, the results obtained with the $V_H$ ($V_D$)
and $V_M$ ($V_G$) potentials coincide.

\subsection{The transition operator}
\label{subsec:transop}

To evaluate the reaction cross section, we need to write down the
nuclear electromagnetic current operator $\mathbf{J}^\dagger(\mathbf{q})$
of Eq.~(\ref{eq:crosssection}). 
This can be written as
\begin{equation}
  \mathbf{J}^\dagger(\mathbf{q})=\int\mathrm{d}\mathbf{x}~
  \mathrm{e}^{i \mathbf{q} \mathbf{x}}~\mathbf{J}(\mathbf{x})
  \label{eq:jq}
\end{equation}
with
\begin{equation}
  \mathbf{J}(\mathbf{x}) = \sum_i ~q_i\,\frac{\mathbf{p}_i}{m_i}~
  \delta^3(\mathbf{x}-\mathbf{x}_i)\:,
\label{eq:jx}
\end{equation}
where $\mathbf{p}_i$, $m_i$, $\mathbf{x}_i$ and $q_i$ are respectively
the momentum,
the mass, the position and the charge of the \textit{i}-th particle.
The matrix element appearing in Eq.~(\ref{eq:crosssection}),
$\hat{\mathbf{\epsilon}}^{\dagger \lambda}_{\mathbf{q}}\cdot \left<\Psi_{^6{\rm Li}}(M)|\mathbf{J}^\dagger(\mathbf{q})|\Psi_{\alpha d}(M_i)\right>$, can be rewritten 
expressing $\Psi_{^6{\rm Li}}(M)$ and $\Psi_{\alpha d}(M_i)$ as
\begin{eqnarray}
  \Psi_{^6{\rm Li}}(M)&=& \frac{\varphi_{0}(r)}{r}Y_{00}(\theta,\phi)\chi_{1 M}+
  \frac{\varphi_{2}(r)}{r}\sum_{m \sigma}\left<2 m, 1 \sigma | 1 M\right>
  Y_{2m}(\theta, \phi)\chi_{1 \sigma}
  \label{eq:psi6}\\
  \Psi_{\alpha d}(M_i)&=& \sum_{\ell_i J_i}\:i^\ell
  \sqrt{4\pi(2\ell_i+1)}\left<\ell_i 0, 1 M_i | J_i M_i \right>\:\nonumber\\
  &\times& \frac{\varphi_{\alpha+d}^{\ell_i J_i}(r)}{kr}\sum_{m' \sigma'}
  \left<\ell_i m', 1 \sigma' | J_i M_i\right>
  Y_{\ell_i m'}(\theta, \phi)\chi_{1 \sigma'}\:,
\label{eq:initialwave}
\end{eqnarray}
where $\varphi_{\ell_f}(r)$ and $\varphi_{\alpha+d}^{\ell_i J_i}(r)$
are the \lith and
\alphad reduced radial functions discussed in Sec.~\ref{subsec:li6-ad}.
In the partial wave decomposition of Eq.~(\ref{eq:initialwave}),
we have retained all the contributions up to $\ell_i=2$.
By then performing a multipole expansion of the
$\mathbf{J}^\dagger(\mathbf{q})$ operator, we obtain
\begin{equation}
\mathbf{\hat{\epsilon}}_\mathbf{q}^{\dagger \lambda}\cdot\mathbf{J}^\dagger(\mathbf{q})=-\sqrt{2\pi}\:\sum_{\Lambda\ge 1}(-i)^\Lambda\sqrt{2\Lambda+1}\left[E_{\Lambda\lambda}(q)+\lambda M_{\Lambda\lambda}(q)\right]\:,
\end{equation}
where $\Lambda$ is the multipole index, while $E_{\Lambda\lambda}(q)$ and
$M_{\Lambda\lambda}(q)$ are the so-called electric and magnetic multipoles of
order $\Lambda$. They are defined as
\begin{eqnarray}
  E_{\Lambda\lambda}(q)   &= &\frac{1}{q}
  \int\mathrm{d}\mathbf{x}\:[\nabla\times(j_\Lambda(qx)
    \mathbf{Y}^\lambda_{\Lambda\Lambda 1}
    (\mathbf{\hat{x}}))]\cdot\mathbf{J}(\mathbf{x})\:,\\
  M_{\Lambda\lambda}(q)   &= &\int\mathrm{d}\mathbf{x}\:j_\Lambda(qx)
  \mathbf{Y}^\lambda_{\Lambda\Lambda 1}
  (\mathbf{\hat{x}})\cdot\mathbf{J}(\mathbf{x})\:,
\end{eqnarray}
where $j_\Lambda(qx)$ is the spherical Bessel function of order $\Lambda$ and $\mathbf{Y}^\lambda_{\Lambda\Lambda 1}(\mathbf{\hat{x}})$ is the vector spherical harmonic of order $\Lambda$.

In this work we adopt the so-called long wavelength approximation (LWA),
since, for the energy range of interest, the momentum of 
the emitted photon is much smaller than the \lith dimension. This means that
we can expand the multipoles in powers of $qr$. Furthermore,
in the present calculation, we include only electric dipole and
quadrupole multipoles, since it has been shown in Ref.~\cite{nol01}
that the magnetic multipoles are expected to give small contributions
to the $S$-factor.

With this approximation, $E_{\Lambda\lambda}(q)$ can be written as
\begin{equation}
  E_{\Lambda\lambda}(q) = Z_e^{(\Lambda)}\:\sqrt{\frac{\Lambda+1}{\Lambda}}
  f_\Lambda(qr)Y_{\Lambda\lambda}(\mathbf{\hat{x}})\:,
\label{eq:electric}
\end{equation}
where~\cite{muk16}
\begin{eqnarray}
f_1(x) &=& 3\frac{[(x^2-2)\sin x+2 x\cos x]}{x^2}\:,\label{eq:f1x}\\
f_2(x) &=& 15\frac{[(5x^2-12)\sin x+(12-x^2) x\cos x]}{x^3}\:,
\label{eq:f2x}
\end{eqnarray}
and $Z_e^{(\Lambda)}$ is the so-called effective charge, and is given
by
\begin{equation}
Z_e^{(\Lambda)}\equiv~Z_d\left(\frac{m_\alpha}{m_\alpha+m_d}\right)^\Lambda +Z_\alpha\left(-\frac{m_d}{m_\alpha+m_d}\right)^\Lambda\:.
\label{eq:zeff}
\end{equation}
Note that when only the first order contribution in the LWA
is retained, $f_\Lambda(x)$ reduces to
\begin{equation}
f_\Lambda(x) = x^\Lambda\:.
\label{eq:flx}
\end{equation}
The use of Eqs.~(\ref{eq:f1x}) and~(\ref{eq:f2x}) instead of Eq.~(\ref{eq:flx})
leads to an increase in the $S$-factor of the order of 1 \%.
This has been shown in Ref.~\cite{muk16} and has been confirmed in the
present work.

\begin{figure}
\centering
\includegraphics[width=1\textwidth]{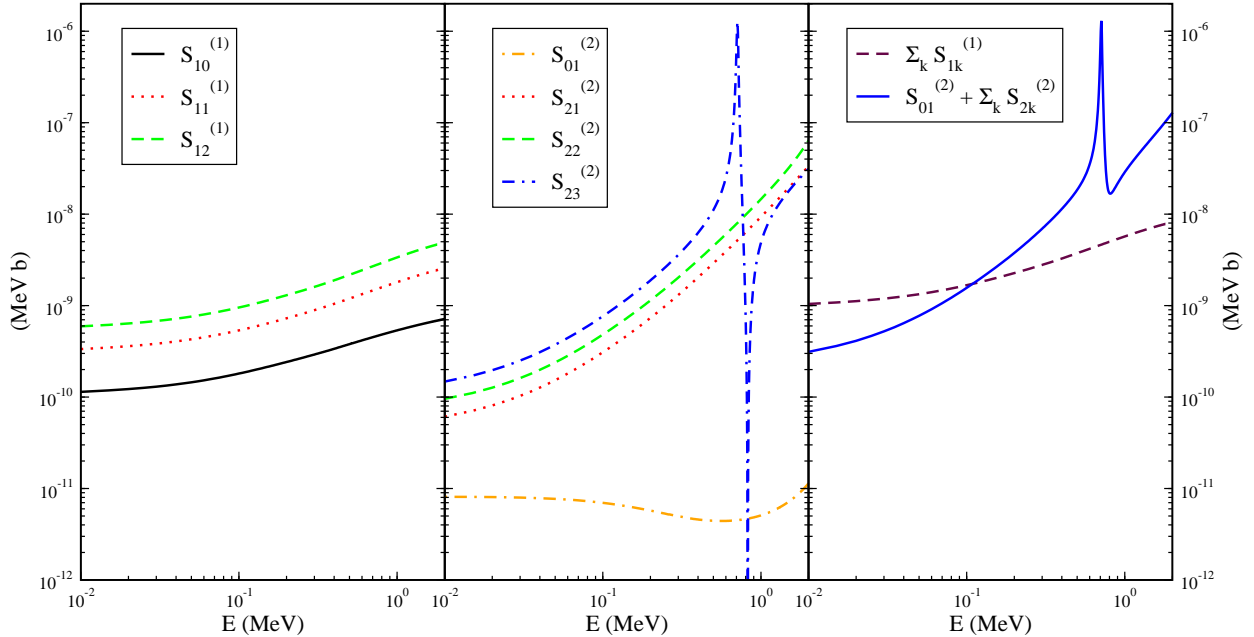}
\caption{The partial astrophysical $S$-factors $S^{(\Lambda)}_{\ell_i J_i}(E)$,
  as defined in Eq.~(\ref{eq:partialSfac}). On the left (central) panel
  the separate contribution for the dipole (quadrupole) transition are shown.
  The shape and color of the lines indicate the initial angular momentum.
  The red dotted and orange dash-dotted lines are used to indicate
  the transitions with $\ell_i=0$ and $\ell_i=2$, respectively, for $J_i=1$.
  For $J_i=0$, $J_i=2$, and $J_i=3$ solid black, green dashed and
  blue dot-dashed-dashed lines are used, respectively. On the right panel the
  total contribution for the dipole (quadrupole) is shown with a maroon
  dashed (blue solid) line.}
\label{fig:sfactorchan}
\end{figure}

In the formalism of the LWA the total cross section of Eq.~(\ref{eq:sigma}) 
can be written as
\begin{equation}
\sigma(E) = \sum_{\ell_i J_i \Lambda}~\sigma_{\ell_iJ_i}^{(\Lambda)}(E)\:,
\label{eq:sigmal}
\end{equation}
where $\sigma_{\ell_iJ_i}^{(\Lambda)}(E)$ is the cross section evaluated with
the electric $\Lambda$-multipole and the initial $\alpha+d$ state
with orbital (total) angular momentum $\ell$ ($J_i$). It can be written as
\begin{multline}
  \sigma_{\ell_iJ_i}^{(\Lambda)}(E) \,=\,\frac{8\pi\:\alpha}{v_{\mathrm{rel}}\,k^2}
  \frac{q}{1+q/m_6}\,\,\frac{Z_e^{(\Lambda)\,2}}{[(2\Lambda+1)!!]^2}
  \frac{(\Lambda+1)(2\Lambda+1)}{\Lambda}
(2\ell_i+1)(2J_i+1)
\\ \times
\bigg[
\sum_{\ell_f}~(-)^{\ell_f}\sqrt{2\ell_f+1}
\begin{pmatrix}
\ell_f &\Lambda&\ell_i\\
0&0&0
\end{pmatrix}
\begin{Bmatrix}
J_i & \ell_i & 1\\
\ell_f & J_f & \Lambda
\end{Bmatrix}\,
\:
\int\:\mathrm{d}r
\:\varphi_{^6{\rm Li}}^{\ell_f}(r)\:f_{\Lambda}(qr)\:\varphi_{\alpha+d}^{\ell_i J_i}(r)
\bigg]^2\:.
\label{crosspartial}
\end{multline}
For simplicity we define the partial $S$-factor as
\begin{equation}
  S^{(\Lambda)}_{\ell_iJ_i}(E) = E\:\sigma^{(\Lambda)}_{\ell_iJ_i}(E)\:
  \exp{\left(2\pi\eta\right)}\:.
\label{eq:partialSfac}
\end{equation}
The results for these quantities evaluated with the $V_G$ potential are
shown in Fig.~\ref{fig:sfactorchan}. The ones for the other potentials have
the same shapes and properties. The only difference comes for \vh, \vt and
\vm, where the contribution to the $S$-factor for the $\ell_i=0$ initial state
are zero, being the transition $^3S_1\rightarrow{^3}S_1$ forbidden for each
multipole term.
Due to the nature of the LWA, the largest contribution to the total
cross section, and therefore to the astrophysical $S$-factor,
should be given by the $E_1$ transition, but, as we can see from
Fig.~\ref{fig:sfactorchan}, the $E_1$ transition dominates only
at energies of the order of few keV. This is due to the so-called $E_1$
isotopic suppression. As we have seen the multipole expansion
at $\Lambda$-th order for the electric terms depends on the
square of the effective charge $Z_e^{(\Lambda)}$ and, for our reaction,
$[Z_e^{(1)}]^2 \simeq 1.6 \times 10^{-5}$ and $[Z_e^{(2)}]^2\simeq 0.44$.
Therefore the $E_1$ contribution to the $S$-factor is strongly suppressed,
except for very low energies,
where the other multipoles are reduced due to their energy dependence.

\subsection{The theoretical astrophysical $S$-factor}
\label{subsec:res}

\begin{figure}[t!]
	\includegraphics[width=1.\textwidth]{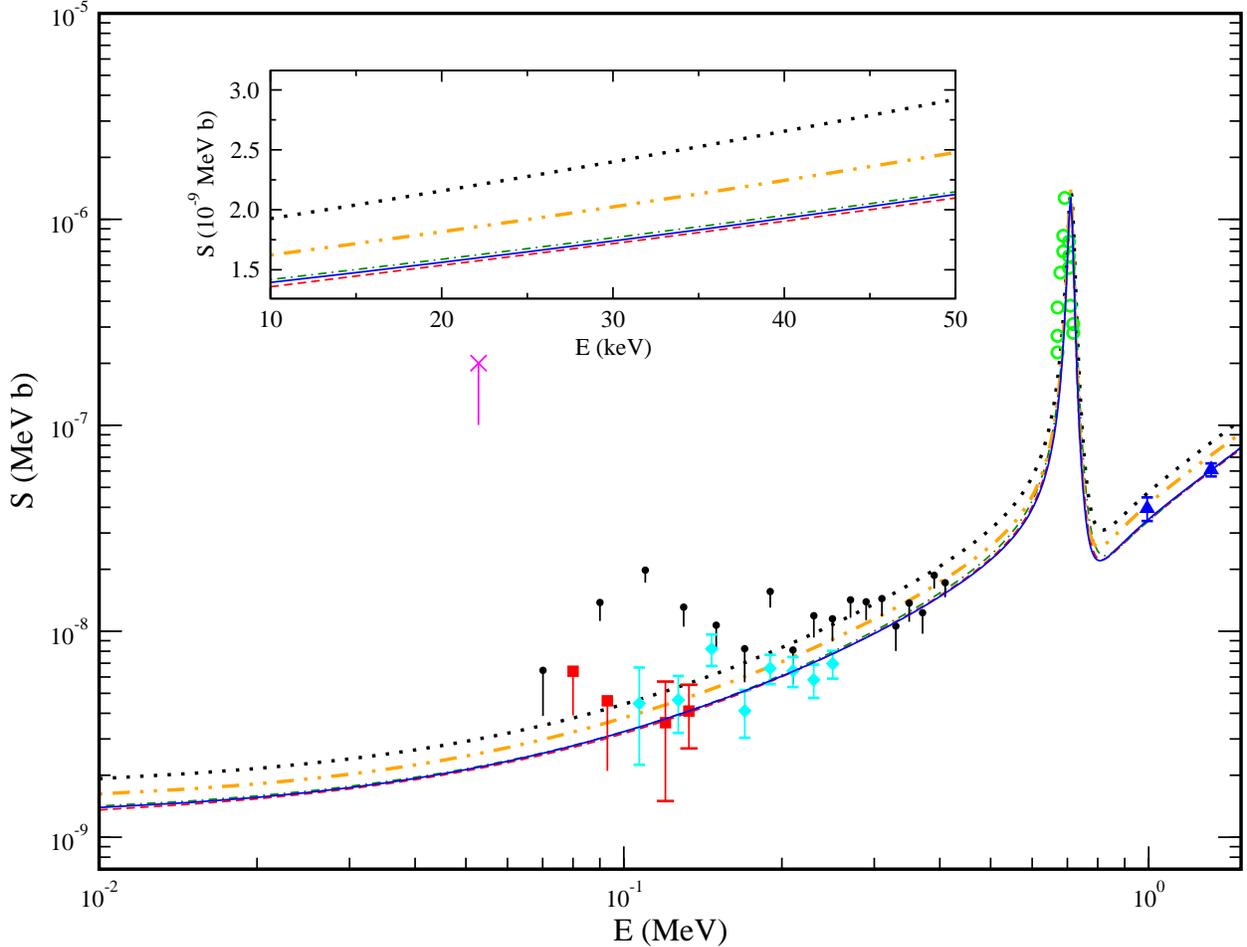}
	\caption{The total astrophysical $S$-factor evaluated with the five
          potential
		models considered in this work is compared with the data
		of Ref.~\cite{rob81} (blue triangles), Ref.~\cite{kie91}
                (black circles),
		Ref.~\cite{moh94} (green circles), Ref.~\cite{cec96}
                (magenta X),
		Ref.~\cite{iga00} (cyan diamonds) and Ref.~\cite{and14,tre17}
                (red squares).
		The data from Refs.~\cite{kie91,cec96} are upper limits
                to the $S$-factor.
		In the insert, the tail of the $S$-factor in the energy range
                10-50
		keV. The dotted (black), dashed (red), dot-dashed (green),
		dot-dot-dashed (orange) and solid (blue) lines correspond
                to the results
		obtained with the $V_H$, $V_T$, $V_M$, $V_D$ and $V_G$
                potentials, respectively.}
	\label{fig:allS}
\end{figure}

\begin{figure}[t!]
	\includegraphics[width=1.\textwidth]{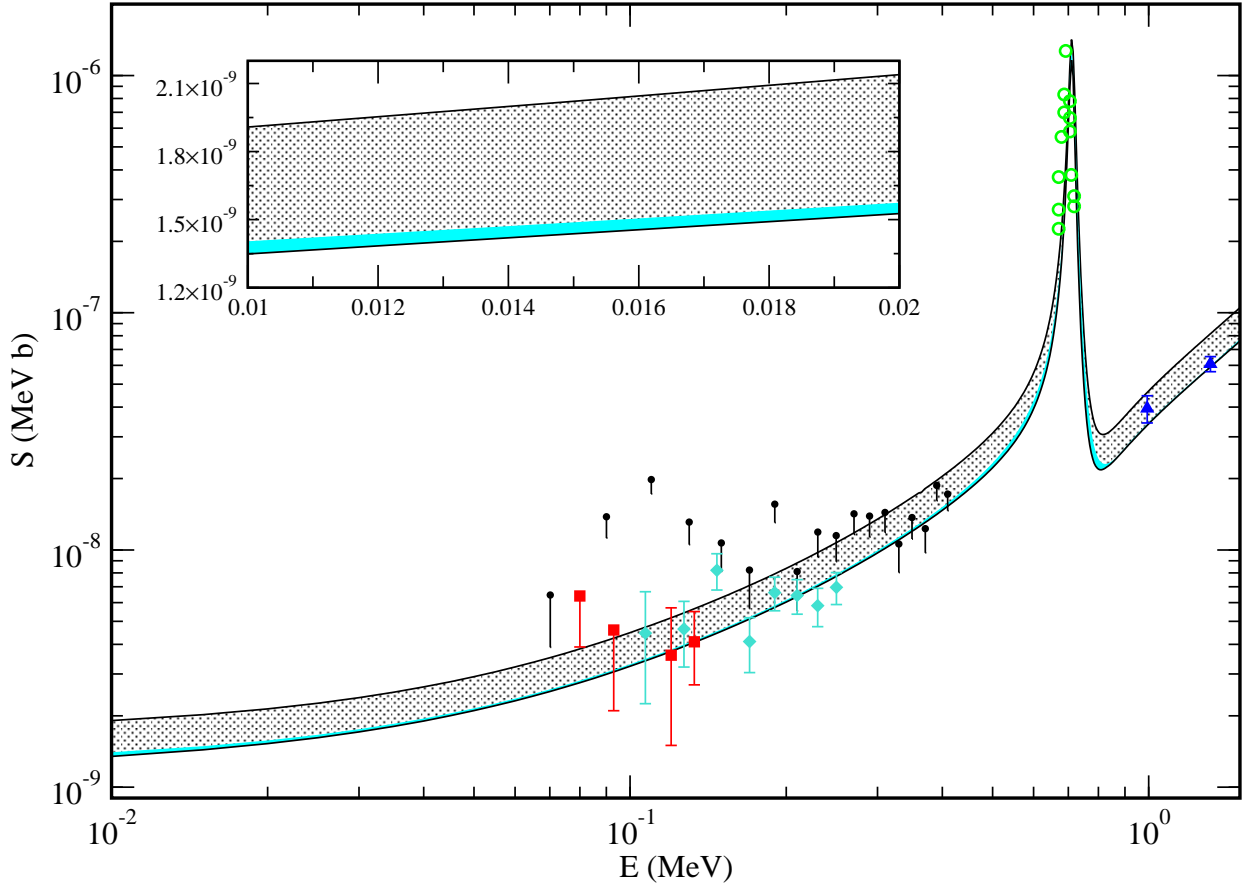}
	\caption{Same as Fig.~\ref{fig:allS} but with the
          theoretical results shown as a band. The (gray) dotted band
          is obtained using all the
          five potentials considered in this work, while the
          narrower (cyan) full band is obtained
          using only those three potentials (\vt, \vm, and \vg)
          which reproduce the experimental
          ANC value.}
	\label{fig:allSerr}
\end{figure}

The calculated astrophysical $S$-factor is compared in Fig.~\ref{fig:allS}
with the available experimental data from
Refs.~\cite{rob81,kie91,moh94,cec96,iga00,and14,tre17}.
By inspection of the figure we can conclude that
the tail of the $S$-factor at low energies
has a strong dependence with respect to the ANC value.
In fact, the three potentials which reproduce the ANC give very close results.
The $V_H$ and $V_D$ potentials, giving a larger value for the ANC
than the other potentials, predict higher values for the $S$-factor.
Thanks to the relatively large number of considered potentials,
we can give a rough estimate of the theoretical uncertainty of our
predictions. Therefore, in Fig.~\ref{fig:allSerr} we show the same results of
Fig.~\ref{fig:allS} as two bands, one obtained using all the five
potentials and a much narrower one calculated with only the three potentials
which reproduce the correct ANC value.
As we can see from the figure, the theoretical uncertainty
for the $S$-factor is much smaller in this second case:
at center-of-mass energies $E\simeq 10$ keV, it is of the order
of 2\%, but it becomes at the 1\% level at the LUNA available energies,
i.e.\ for $E\simeq 100$ keV.
On the other hand, if we consider all of the potentials, the previous
estimates grow to 25\% and 24\% at $E\simeq 10$ and 100 keV, respectively.
The available experimental data, though, are not
accurate 
enough in order to discriminate between the results obtained
with these five potentials.
Therefore, in the following Section, where the primordial \lith
abundance is discussed, we consider conservatively the results for the
astrophysical $S$-factor obtained with all the five potentials.

\section{The \lith primordial abundance}
\label{sec:li6PA}

$^6$Li is expected to be produced during BBN with a rather low number
density, $^6$Li/H $\sim 10^{-14}$, for the baryon density as obtained by the
2015 Planck results \cite{Ade:2015xua}. This
result still holds using the $S$-factor described in the previous
Section (see below), and it is too small to be detectable at present. 
Actually, some positive measurements in old halo stars at the level of
$^6$Li/$^7$Li $\simeq 0.05$ were obtained in the last decade~\cite{asp06},
but they may reflect the post-primordial production of this nuclide in Cosmic
Ray spallation nucleosynthesis. Moreover, as we mentioned already, a more
precise treatment of stellar atmosphere, including convection, shows that
stellar convective motions can generate asymmetries in the line shape that
mimic the presence of $^6$Li, so that the value 0.05 should be rather
understood as a robust upper limit on $^6$Li primordial abundance. This does
not mean that the issue is irrelevant for BBN studies since  the study of the
chemical evolution of the fragile isotopes of Li, Be and B could constraint
the $^7$Li primordial abundance, and clarify the observational situation of
Spite Plateau, see e.g. Ref.~\cite{Olive:2016xmw}.

\begin{figure*}
\begin{center}
\includegraphics[width=.9\textwidth]{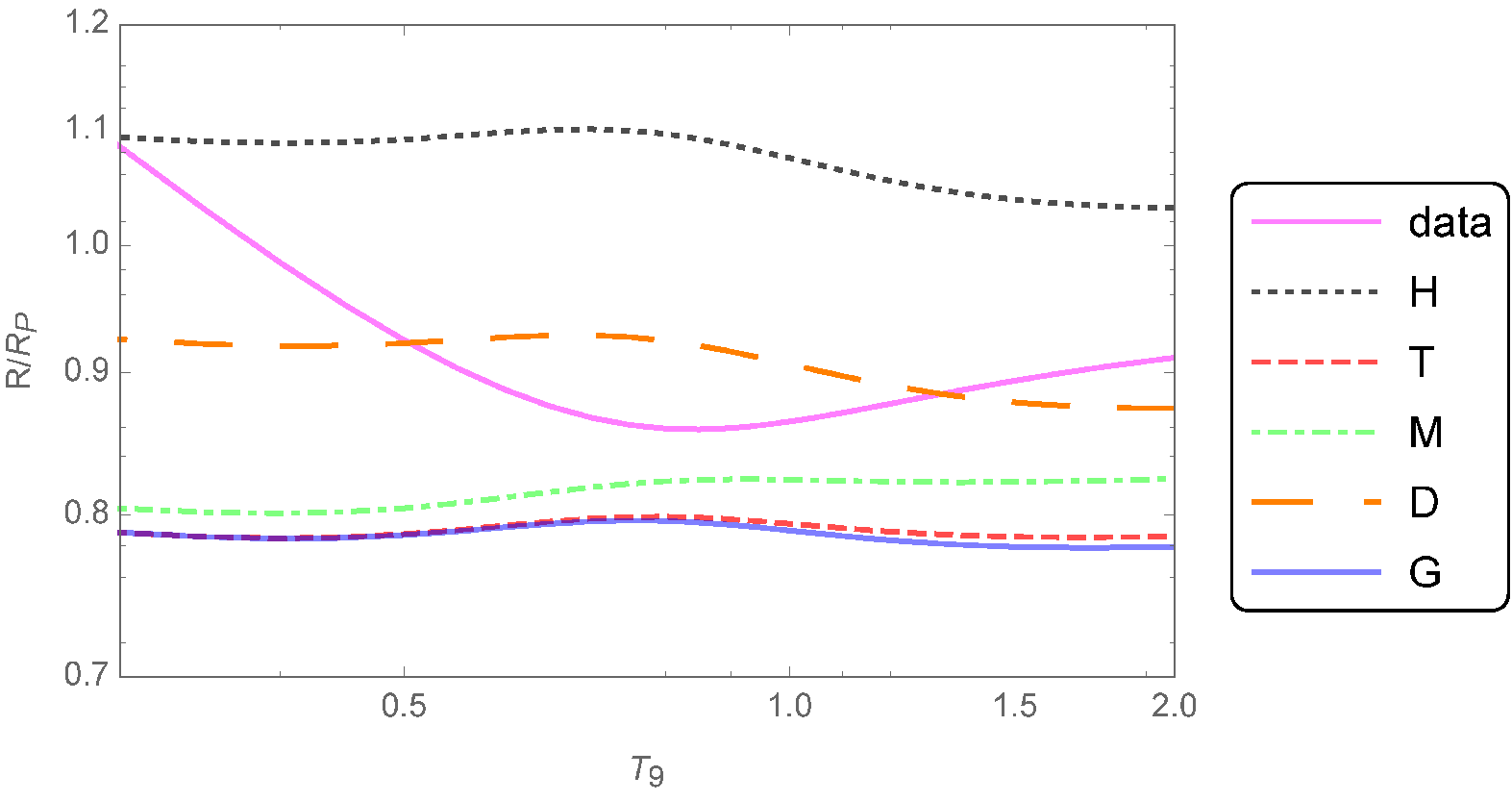} 
\end{center}
\caption{Rates vs. the temperature $T$ in units of $10^9$ K ($T_9$),
  corresponding to the astrophysical $S$-factors of the data
  fit (solid/magenta), and of the theoretical calculations with the
  five potentials used in this work
  (dotted/black, dashed/red, dot-dashed/green, long-dashed/orange and
  solid/blue, corresponding to $V_H$, $V_T$, $V_M$, $V_D$ and $V_G$ potentials,
  respectively), normalized to the standard rate used in PArthENoPE
  (NACRE 1999). \label{f:rates}} 
\end{figure*}

The whole $^6$Li is basically produced via the $\alpha+d$ process,
which is thus the leading reaction affecting the final yield of this isotope. 
The new theoretical  $S$-factors detailed so far have been used
to compute the thermal rate in the BBN temperature range, by folding the cross
section with the Maxwell-Boltzmann distribution of involved nuclides. We have
then changed the PArthENoPE code~\cite{Pisanti:2007hk} accordingly, and
analyzed the effect of each different $S$-factor on the final abundance of
$^6$Li, as function of the baryon density. For comparison, we also consider
the value of the $S$-factor as obtained from fitting experimental data 
from Refs.~\cite{RO81,MO94,IG00,AN14}
 \bea
 S(E) &=& 10^{-9} \left( 3.19368 + 6.94243\, E + 32.204\, E^2 \right) 
 \nonumber \\
&+& \frac{9.96936 \times 10^{-7}}{1. + 4800.46 \,(E - 0.694061)^2} \ ,
\label{e:astro}
\eea
as well as the NACRE 1999 fit~\cite{nacre99},
which is used as benchmark rate in PArthENoPE public code.
The results are shown in Fig.~\ref{f:rates}, normalized to NACRE 1999.
As we can see, the change is in the 10-20 \% range. If we adopt the
Planck 2015 best fit for the baryon density parameter
$\Omega_b h^2= 0.00226$~\cite{Ade:2015xua}, we obtain values for the
$^6$Li/H density ratio in the range $(0.9 - 1.4)\times 10^{-14}$, slightly
smaller than what would be the result if the experimental data fit
is used, as it can be seen in Table~\ref{op}.
\begin{table}[b]
\centering
\begin{tabular}{|c|c|c|c|c|c|c|c|} \hline
  & bench & ~data~ & ~~H~~ & ~~T~~ & ~~M~~ & ~~D~~ & ~~G~~ \\
  \hline
  $^6$Li/H $\times 10^{14}$ & $1.1$ & $1.7$
  & $1.4$ & $1.1$ & $1.1$ & $1.0$ & $0.93$ \\ \hline
\end{tabular}
\caption{Values of the final yield of $^6$Li (relative to H) for the
  five potential models considered in this paper, as well as for the
  NACRE 1999 rate, used as benchmark in PArthENoPE (bench)
  and using a fit of experimental data (data).}
\label{op}
\end{table}

\begin{figure*}
\begin{center}
\includegraphics[width=.9\textwidth]{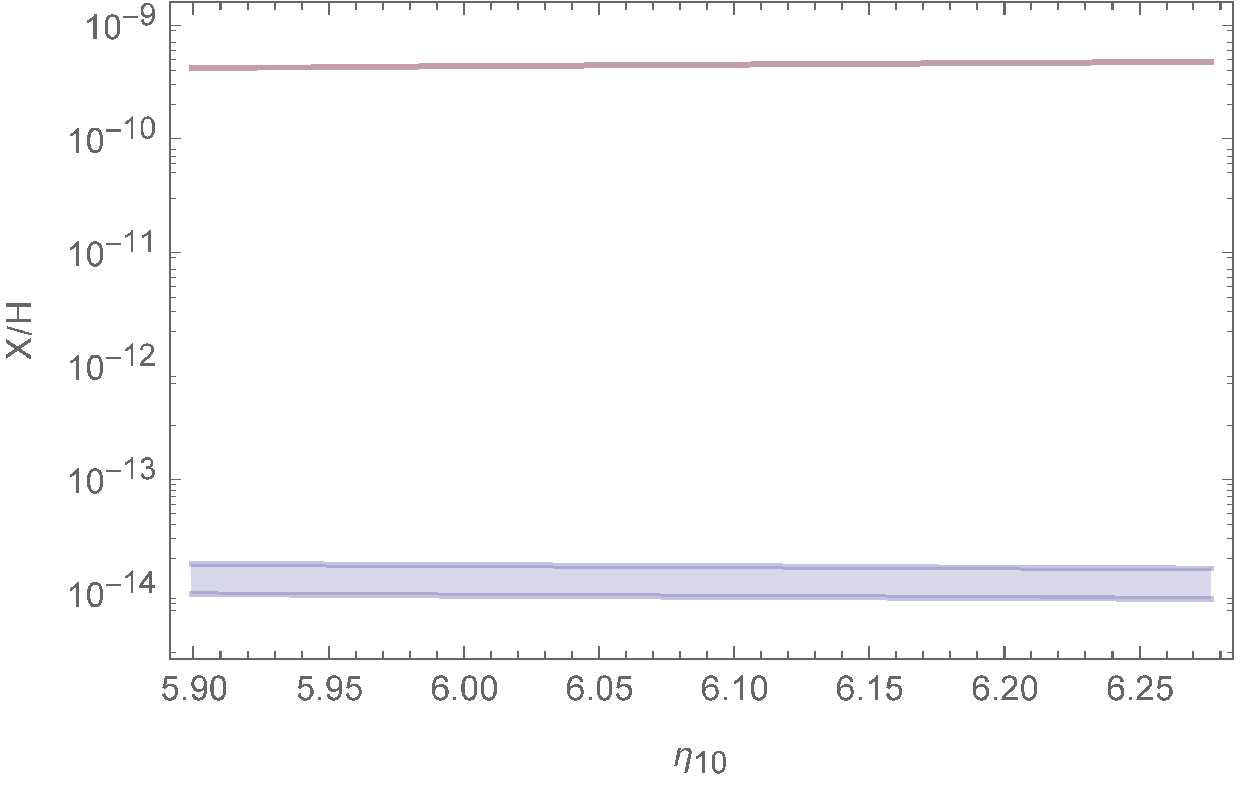} 
\end{center}
\caption{The X/H abundance for X=$^6$Li (lower band) and X= $^7$Be+$^7$Li
  (upper band). The theoretical uncertainty arising from the use of
  the five potential models considered in this paper is shown as a band.
  \label{f:Li6}} 
\end{figure*}

Notice that, at least with present sensitivity on $^6$Li yields, the
dependence on the baryon density, or equivalently, the baryon to photon
density ratio $\eta_{10} \sim 273.49 \, \Omega_b h^2$, is quite mild, as
shown in Fig.~\ref{f:Li6}. The lower band in this plot cover the range of
values obtained when the five potential models are used,
and we can conservatively say
that standard BBN predicts $^6$Li/H$ = (0.9 - 1.8) \times 10^{-14}$. This
range is also in good agreement with the results of other
studies~\cite{ham10,Cyburt:2015mya}.  In Fig.~\ref{f:Li6} we also show the
final abundance of $^7$Be+$^7$Li (upper band), which remains
in the range $(4.2 - 4.7) \times 10^{-10}$,
and it is, as expected, almost independent of the potential model adopted
for the $\alpha+d$ radiative capture reaction considered here.
\section{Conclusions}
\label{sec:concl}
The $\alpha+d$ radiative capture has been studied within a two-body
framework, where the $\alpha$ particle and the deuteron are considered
as structureless constituent of $^6$Li. The long-wavelength approximation (LWA)
has been used, and the electric $E_1$ and $E_2$ multipoles have been retained.
In order to study the accuracy that the present theoretical framework can reach,
we have used five different models for the $\alpha+d$ interaction, among which
also, for the first time, potential models with a tensor term, able to
reproduce the magnetic dipole and electric quadrupole moments of \lith, as
well the $S$-state ANC and the $\alpha+d$ scattering phase shifts. The
theoretical uncertainty to the astrophysical $S$-factor, the observable
of interest, is of the order of $\sim$ 20\% if all the five potential
models are retained, but reduces to few \% if only those potentials
which reproduce the $S$-state ANC are considered. The experimental data,
however, are affectd by an uncertainty much larger than the theoretical one.

The calculated values for the $\alpha+d$ astrophysical $S$-factor
have been used in the PArthENoPE public code in order to estimate the
\lith and $^7$Li+$^7$Be primordial abundances. The \lith
abundance is predicted to be slightly smaller
than what would result from the available experimental data and from
the NACRE 1999 compilation,
but still in the range of $(0.9 - 1.8) \times 10^{-14}$.
We conclude that this result of standard BBN is thus quite robust.
Further studies about $^6$Li astrophysical measurement may be needed
to check the claim of a much larger ratio $^6$Li/$^7$Li obtained in
Ref.~\cite{asp06}. On the other hand, the
final $^7$Li+$^7$Be abundance is almost independent on the result for
the astrophysical $S$-factor presented here, and is found to be
in the range of $(4.2 - 4.7) \times 10^{-10}$.

Finally, we would like to notice that 
the present calculation for the astrophysical $S$-factor
is, to our knowledge,
the most up-to-date one working within
a two-body framework. However, the assumption that the deuteron is
a structureless constituent of \lith can be considered rather weak,
and the present study could be improved if
the six-body systems are viewed as a core of an $\alpha$ particle
and two nucleons, i.e.\ as a three-body systems. The first steps within
this three-body framework have been done in Ref.~\cite{Tur16},
and further work along this line is currently underway.

\end{document}